\documentclass[12pt,preprint]{aastex}

\shorttitle{THE MULTIPLE POPULATION IN NGC 288}
\shortauthors{Roh et al.}

\begin{document}

\title{TWO DISTINCT RED GIANT BRANCHES IN THE GLOBULAR CLUSTER NGC 288}

\author{DONG-GOO ROH\altaffilmark{1}, YOUNG-WOOK LEE\altaffilmark{1},
        SEOK-JOO JOO\altaffilmark{1}, SANG-IL HAN\altaffilmark{1},
        YOUNG-JONG SOHN\altaffilmark{1}, AND JAE-WOO LEE\altaffilmark{2} }
\affil{\altaffilmark{1}{Center for Galaxy Evolution Reseach and Department of Astronomy, Yonsei University, Seoul 120-749, Korea} {; ywlee2@yonsei.ac.kr}\\
\altaffilmark{2}{Department of Astronomy and Space Science, Sejong University, Seoul 143-747, Korea}}

\begin{abstract}

We report the presence of two distinct red giant branches (RGBs) in the globular cluster 
NGC 288 from the narrow-band calcium and Str\"{o}mgren $b$ \& $y$ photometry obtained at the CTIO 4m Blanco telescope.
The RGB of NGC 288 is clearly split into two in the $hk$ [=$(Ca - b) - (b - y)$] index, 
while the split is not shown in the $b - y$ color. 
Unlike other globular clusters with multiple populations reported thus far, the horizontal branch of NGC 288 is only mildly extended.
Our stellar population models show that this and the presence of two distinct RGBs in NGC 288
can be reproduced if slightly metal-rich ($\Delta [m/H]$ $\approx$ 0.16) second generation stars are 
also enhanced in helium by small amount ($\Delta Y$ $\approx$ 0.03) and younger by $\sim$ 1.5 Gyrs.
The RGB split in $hk$ index is most likely indicating that the second generation stars were affected 
by supernovae enrichment, together with the pollutions of lighter elements 
by intermediate-mass asymptotic giant branch stars or fast-rotating massive stars.
In order to confirm this, however, spectroscopy of stars in the two distinct RGB groups is urgently required.
\end{abstract}


\section{INTRODUCTION}

Unlike the conventional wisdom, observations made during the past decade have revealed that 
many globular clusters (GCs) are possessing more than one stellar population.
Some of these peculiar GCs, such as $\omega$~Cen \citep{Lee99,Bedin04}, M54 \citep{Layden97,Siegel07}, 
M22 \citep{DaCosta09,Lee09a,Marino09}, NGC 1851 \citep{Han09,Lee09b,Carretta10}, Terzan~5 \citep{Ferraro09}, 
and NGC 2419 \citep{Cohen10,DiCriscienzo11} show evidences of supernovae (SNe) enrichment, 
indicating that they are relics of more massive primeval dwarf galaxies, rather than being normal GCs.
For other GCs with multiple populations, such as NGC 2808 \citep{Piotto07}, NGC 6388 \citep{Moretti09}, 
and M4 \citep{Marino08}, however, the evidence for the discrete distribution of heavy elements 
as observed in the RGB of $\omega$~Cen is lacking, although spreads in some lighter elements \citep[][and references therein]{Carretta09} and helium \citep{DAntona05,Lee05,Piotto07,Yoon08} are reported.
Therefore, the presence of chemical inhomogeneity and multiple populations in these GCs is 
largely considered due to the pollution from the intermediate-mass asymptotic giant branch (AGB) stars 
and (or) fast-rotating massive stars \citep{Ventura08,Decressin07}, which is expected even in normal GCs.

Because of their important implications on the hierarchical merging paradigm of Galaxy formation, 
search for more GCs with dwarf galaxy origin (i.e., with evidence of SNe enrichment) would be extremely important.
The purpose of this Letter is to report that 
NGC 288 is also showing a clear split in the RGB from the narrow-band calcium photometry.
This observation is compared with our stellar population models to argue that 
the two populatons are different in terms of overall metallicity, helium, and age by small amounts.

\section{OBSERVATIONS AND COLOR-MAGNITUDE DIAGRAMS}

Our observations in $Ca$, $b$, and $y$ passbands were performed using the CTIO 4m Blanco telescope on 2009 July 27.
The telescope was equipped with eight 2048$\times$4096 SITe CCDs, providing a plate scale of 0.27 arcsec pixel$^{-1}$ and a field-of-view of 36$\times$36 arcmin on the sky.
However, since the 4$\times$4 inch filters used in our photometry can not cover the entire field, 
only four CCD chips located in central region (i.e., chip 2, 3, 6, and 7) covering 2048$\times$2500 pixels per chip were used in the final photometry.
The total exposure times for $Ca$, Str\"{o}mgren $b$, and $y$ were 1650, 264, and 132 seconds, 
respectively, split into short and long exposures in each band.
NGC 288 was placed on chip 6, approximately 3.0 arcmin South and 3.1 arcmin East from the CCD center.
The IRAF\footnote{IRAF is distributed by the National Optical Astronomy Observatories, which are operated by the Association of Universities for Research in Astronomy, Inc., under cooperative agreement with the National Science Foundation.} MSCRED Package was used for preprocessing including bias correction and flat fielding.
The brightnesses of objects in NGC 288 were measured with
the point-spread function (PSF) fitting routine DAOPHOT II and ALLFRAME \citep{Stetson87,Stetson94},
and aperture corrections were calculated using the DAOGROW \citep{Stetson90}.
Our photometry in $Ca$, $b$, and $y$ passbands were then used to calculate the $hk$ $[= (Ca-b)-(b-y)]$ index defined by \citet{Anthony91}.
The $Ca$ filter in the $hk$ index is meant to measure essentially ionized calcium H and K lines, 
and the $hk$ index is known to be about three times more sensitive to metallicity than 
$m_{1}$ $[= (v-b)-(b-y)]$ index is \citep{Twarog95}. 
The same filter set employed in this observation was extensively used by us in our previous investigations of GCs \citep{Rey00,Rey04,Lee09a,Lee09b}.

Figure 1 shows color-magnitude diagrams (CMDs) of NGC 288 
in ($b-y$, $y$) and ($hk$, $y$) planes.
To examine the CMD features more clearly, magnitude error, chi, sharpness and separation index \citep{Stetson03} were used to reject stars with large photometric uncertainty and those affected by blending and adjacent starlight contamination.
All stars in the Figure 1 lie within the chip 6, 
and therefore our CMDs are not subject to any uncertainty stemming from the possible chip to chip variations of the mosaic CCDs.
The most remarkable feature of Figure 1 is the presence of two distinct RGBs in the $hk$ vs.\ $y$ CMD.
When measured at $y = 16.5$ mag, 
the mean separation between the two RGBs is about 0.10 mag in $hk$ index.
The discrete distribution shown in RGB, however, is not apparent in the subgiant branch (SGB).
Note also that the RGB split is not shown in ($b-y$, $y$) CMD.
This is most likely because the $Ca$ filter in the $hk$ index is much more sensitive to changes in $Ca$ abundance than other color indices like $b-y$.

Given the small foreground reddening value of $E(B-V)$ = 0.03 \citep{Harris96} toward NGC 288, 
it is very unlikely that the differential reddening has caused the double RGBs.
Furthermore, in contrast to other color indices, the $hk$ index is known to be insensitive 
to interstellar reddeing, $E(hk)/E(B-V)$ = $-$0.12 and $E(hk)/E(b-y)$ = $-$0.16 \citep{Anthony91}.
Therefore, if we adopt the reddening of $E(B-V)$ = 0.03, $E(hk)$ = $-$0.0036 is obtained for NGC 288,
which is negligible compared to the separation in $hk$ index ($\sim$ 0.10 mag) between the two RGBs. 
Star counts of two subpopulations indicate that  
the bluer RGB population (``Pop-1'') takes about 60\% of the whole population, 
while the redder RGB population (``Pop-2'') comprises about 40\% of total population, 
in the magnitude interval $y = 16.5 \pm 1.5$ mag.
This ratio is not sensitive to the adopted separation index in our photometry.

\section{COMPARISON WITH STELLAR POPULATION MODELS}

In order to better understand the origin of the RGB split in $hk$ index, and to place constraints on the chemical combinations of two subpopulations,
we have constructed stellar population models based on the latest version of the Yonsei-Yale (Y$^{2}$) isochrones \citep{Yi08}  
and HB evolutionary tracks (S.-I. Han et al. 2011, in preparation).
Readers are referred to \citet{Lee90,Lee94} and \citet{Yoon08} for the details of our model construction.
Figures 2 and 3 present our synthetic CMDs for NGC 288
in ($hk$, $y$) and ($b-y$, $y$) planes, respectively.
Our models were constructed under three different assumptions regarding the chemical enrichment and age spread in NGC 288.
First, we assumed that the second generation population (Pop-2; redder RGB) is more enhanced in metallicity and younger ($\Delta [m/H]$ $\approx$ 0.16 dex, $\Delta t$ $\approx$ 1.5 Gyr),
but not enhanced in helium abundance [hereafter $\Delta$Z+$\Delta$Age model; panel (b) in Figures 2 and 3].
These models match well with the observed CMDs from the MS through the RGB in ($hk$, $y$) and ($b-y$, $y$) planes.
Yet, the models fail to reproduce the HB, as the synthetic HBs are too extended in color 
including significant numbers of RR Lyraes and red HB stars (HB type\footnote{The HB type is the quantity, (B$-$R)$/$(B$+$V$+$R), where B, V, and R are the numbers of blue HB, RR Lyrae variable, and red HB stars, respectively \citep{Lee94}.}  = 0.61), 
while the observed HB is only mildly extended with mostly blue HB stars (HB type = 0.91).
This is because both metal enhancement and younger age in second population move the HB to red in CMD \citep[see][]{Lee94}.
Second, we then assumed that both metal and helium abundances are enhanced ($\Delta [m/H]$ $\approx$ 0.16 dex, $\Delta Y$ $\approx$ 0.03 dex), 
but age is constant [hereafter $\Delta$Z+$\Delta$Y model; panel (c) in Figures 2 and 3].
These models match well with the observed CMDs from the RGB through the HB in ($hk$, $y$) and ($b-y$, $y$) planes.
However, they can not reproduce the narrow SGB in ($hk$, $y$) CMD.
Therefore, both $\Delta$Z+$\Delta$Age and $\Delta$Z+$\Delta$Y models are in conflict with the observed CMDs of NGC 288.

Finally, we assumed that not only metal and helium abundances are enhanced, 
but also age is younger in Pop-2 [hereafter $\Delta$Z+$\Delta$Y+$\Delta$Age model; panel (d) in Figures 2 and 3].
These models are in good agreements with the observations from the MS to the HB.
The enhanced metal abundance in Pop-2 makes the RGB split in $hk$ index as observed, 
while the younger age can explain the narrow and apparently single SGB in ($hk$, $y$) CMD.
The increase in helium abundance in Pop-2 moves HB bluer \citep[see][]{Lee05}, 
almost cancelling out the effects by enhanced metallicity and younger age, 
making the blue HB only mildly extended with the HB type similar to the observed value (HB type = 0.90).
Note that, in our HB simulations, we employ the standard \citet{Reimers77} mass-loss law and the same mass-loss parameter $\eta$ for the two subpopulations. 
Input parameters adopted in our best models (i.e., $\Delta$Z+$\Delta$Y+$\Delta$Age model) are listed in Table 1.
Our models are computed with the same abundance of [CNONa/Fe] for both Pop-1 and Pop-2.
More detailed models including the possible difference in [CNONa/Fe] between the two subpopulations would change the age estimates \citep[see, e.g.,][]{Cassisi08}.

\section{DISCUSSION}

We have shown that the RGB of NGC 288 is split into two distinct sequences.
While this is most likely the effect of Ca II H \& K lines, 
it is important to check whether the CN band at $\lambda$~$\approx$ 3870 \AA $ $ could affect the $hk$ index 
due to the proximity of the CN band to the blue tail of $Ca$ filter transmission curve.
In particular, according to \citet{Kayser08}, lower RGB stars in NGC 288 exhibit CN bimordality that spans about 0.6 dex.
In Figure 4a, we have matched their spectroscopic data with our photometry,
where we can see that ``CN-stong'' stars lie well on the redder RGB sequence, 
whereas ``CN-weak'' stars are on the bluer RGB.
This further suggests that possible contamination of CN band to $hk$ index should be investigated in more detail.
It is known that the passband of the narrow band interference filter, such as the $Ca$ filter employed in our photometry, depends on the angle of incidence beam \citep[see][]{Clarke75,Lee09a}.
This issue is therefore more relevant when the $Ca$ filter is used with a relatively fast telescope like the prime focus of the CTIO 4m telescope,
where the central wavelength drift of the $Ca$ filter is estimated to be about 15 \AA $ $ to the shorter wavelength\footnote{http://www.ctio.noao.edu/instruments/filters/filters\_66.html} (see Figure 4b).

In order to see the influence of CN band on the $hk$ index in our photometry, 
we have calculated synthetic spectra using the ATLAS 9 model atmosphere \citep{Castelli03} for the RGB star 
at the magnitude level of HB with $T_{\rm eff}$~=~$4750$~K, \(\log g=\)~$2.0$ (in cgs unit), 
$v_{\rm turb}$~=~$2.0$~km/s, and [Fe/H]~=~$-1.6$ (see Lee et al. 2009a Supplementary Information for detail).
The typical RGB stars in GCs show an anticorrelation between CN band and CH band strengths and a correlation between CN band and NH band strengths, 
indicating that the nitrogen controls the CN band strength \citep{Briley93}.
Therefore, we have compared the synthetic spectra between the stars with normal and enhanced nitrogen abundances (see Figure 4c).
We obtain $\Delta hk$ $\approx$ $+$0.006 for $\Delta$ [N/Fe] = $+$1.0 dex, 
which suggests that the influence of CN band on the $hk$ index would be negligible 
in our photometry\footnote{If we have assumed the wavelength drift of 20 \AA, a similarly small value of $\Delta hk$ ($\sim$ $+$0.009) is obtained, confirming that this result is not very sensitive to the uncertainty in the value of the wavelength drift.}.
Recently, \citet{Sbordone11} also reached at a similar conclusion, 
finding that CNONa anticorrelations have an effect of at most $\sim$ 0.04 mag in the $hk$ color at fixed $M_{v}$.
Their calculations include not only the direct effects of CN and other absorption bands (NH \& CH), 
but also the effects on the continuum levels in the $Ca$, $b$, and $y$ filters.

If the CNONa abundances have only little effects on the $hk$ index as discussed above, which should be confirmed in the forthcoming works,
the RGB split discovered in our $hk$ vs.\ $y$ CMD would indicate a small difference in $Ca$ abundance between the two subpopulations.
Since calcium and other heavy elements can only be supplied through SNe explosions, 
this in turn would suggest that the second generation stars were affected by SNe enrichment,
together with the pollutions of lighter elements (such as the enhancenment of N and the depletion of O) 
by intermediate-mass asymptotic giant branch stars or fast-rotating massive stars.
Spectroscopy of stars in the two distinct RGB sequences is
crucial to confirm the small difference in the abundance of heavier elements suggested in our photometry.

\acknowledgments{We thank the referee for a number of helpful suggestions. Support for this work was provided by the National Research Foundation of Korea to the Center for Galaxy Evolution Research. This material is based upon work supported by AURA through the NSF under AURA Cooperative Agreement AST 0132798, as amended.}

\clearpage
\begin{figure*}[ht]
\epsscale{0.9}
\plotone{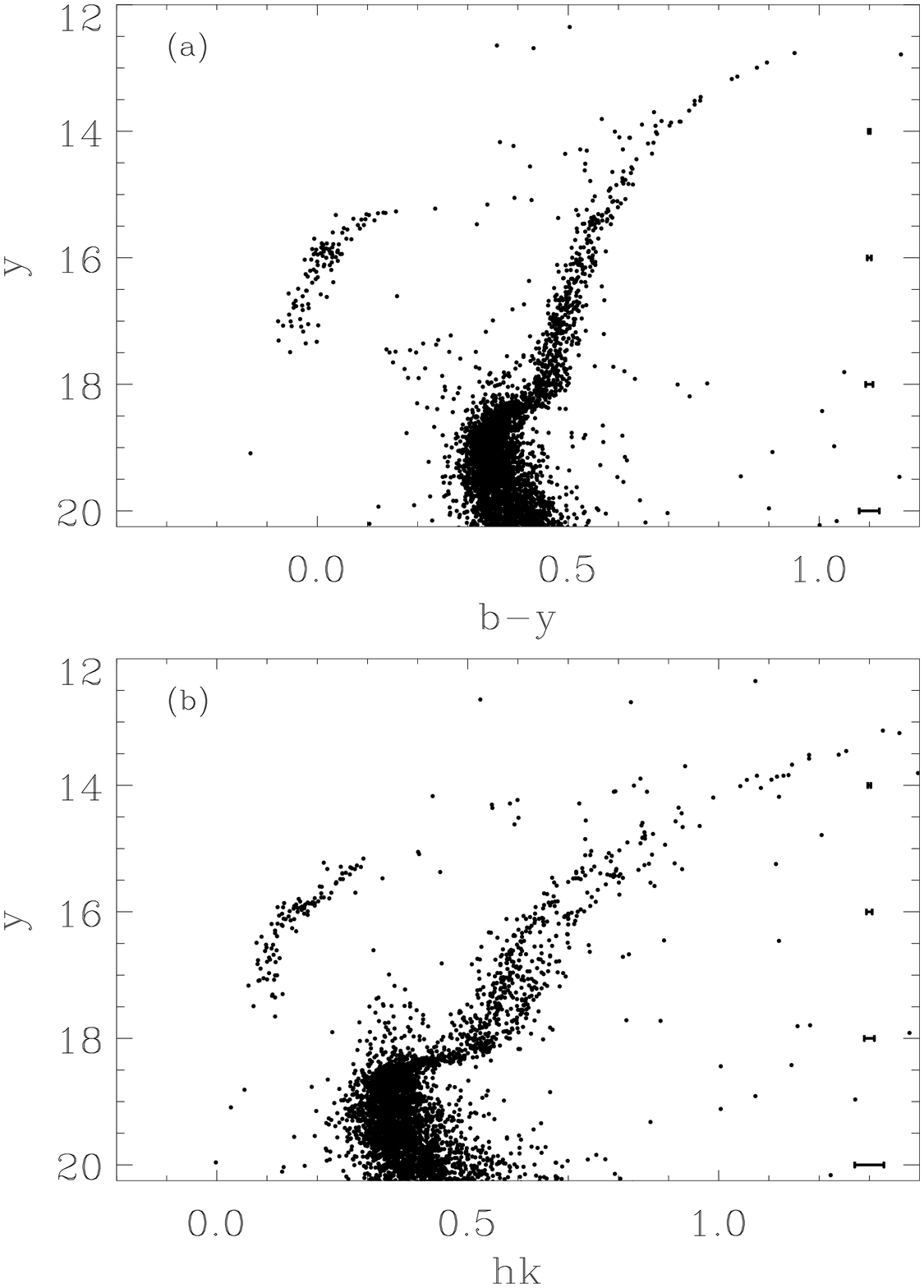}
\caption{Color-magnitude diagrams for NGC 288 from CTIO 4m Blanco Telescope. Note the discrete double RGBs in the $hk$ vs.\ $y$ CMD.
Two RGBs are separated by $\sim$ 0.10 mag in $hk$ index.
The reddening vector values are too small [$E(hk)$ = $-$0.0036, $E(b-y)$ = 0.0225, and $A_{y}$ = 0.093] to present in this figure.
\label{fig1}}
\end{figure*}

\clearpage
\begin{figure*}[ht]
\epsscale{0.9}
\plotone{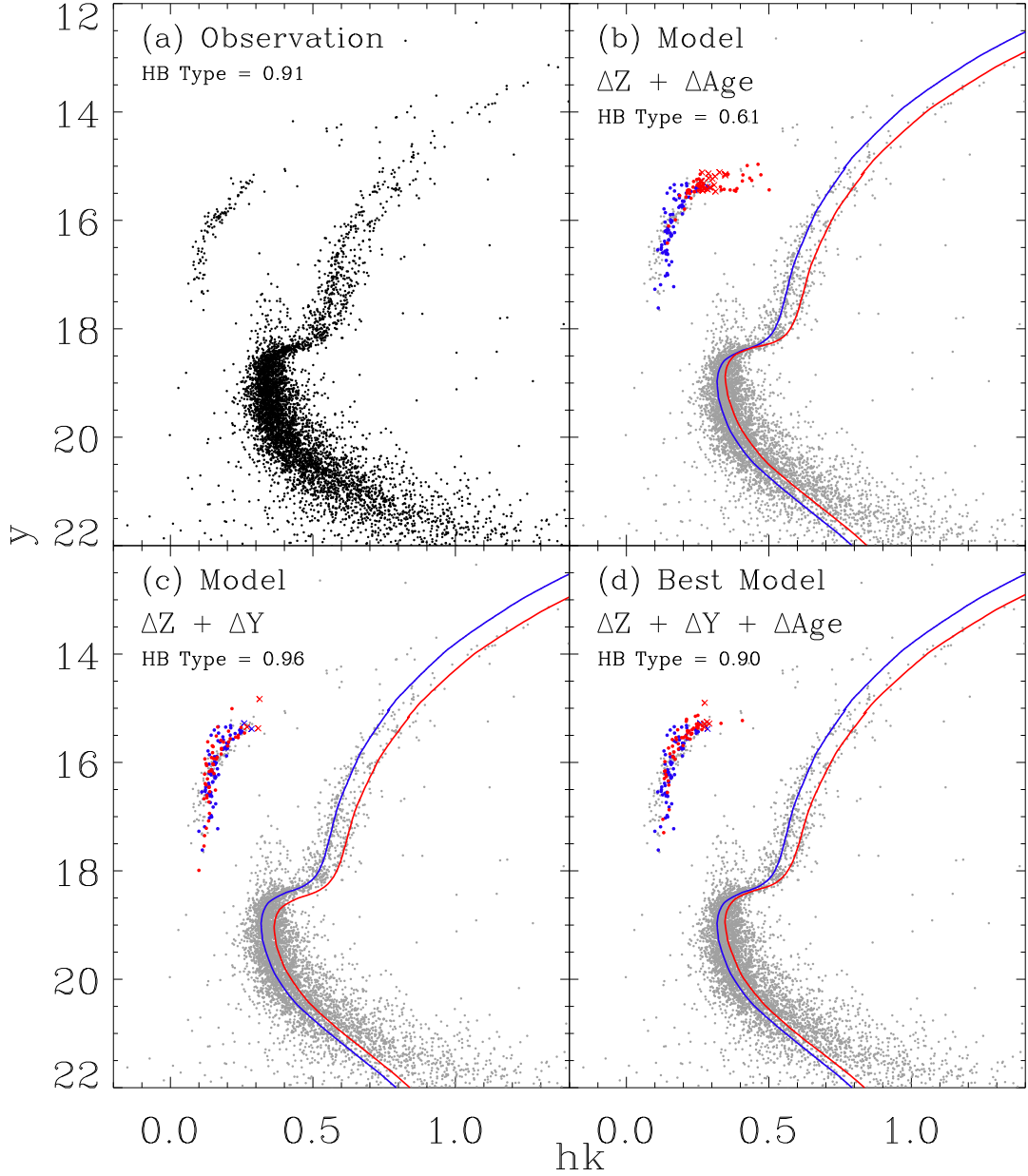}
\caption{Our population models for NGC 288 compared with the observed CMD in $hk$ vs.\ $y$ plane.
In the models, the red lines and symbols are for the metal-rich second generation population,
while the blue lines and symbols are for the metal-poor poupulation.
The crosses in the models denote RR Lyrae variables.
The best match is obtained when metal-rich ($\Delta [m/H]$ $\approx$ 0.16) second generation stars are
also enhanced in helium by small amount ($\Delta Y$ $\approx$ 0.03) and younger by $\sim$ 1.5 Gyrs (see text).
\label{fig2}}
\end{figure*}

\clearpage
\begin{figure*}[ht]
\epsscale{0.9}
\plotone{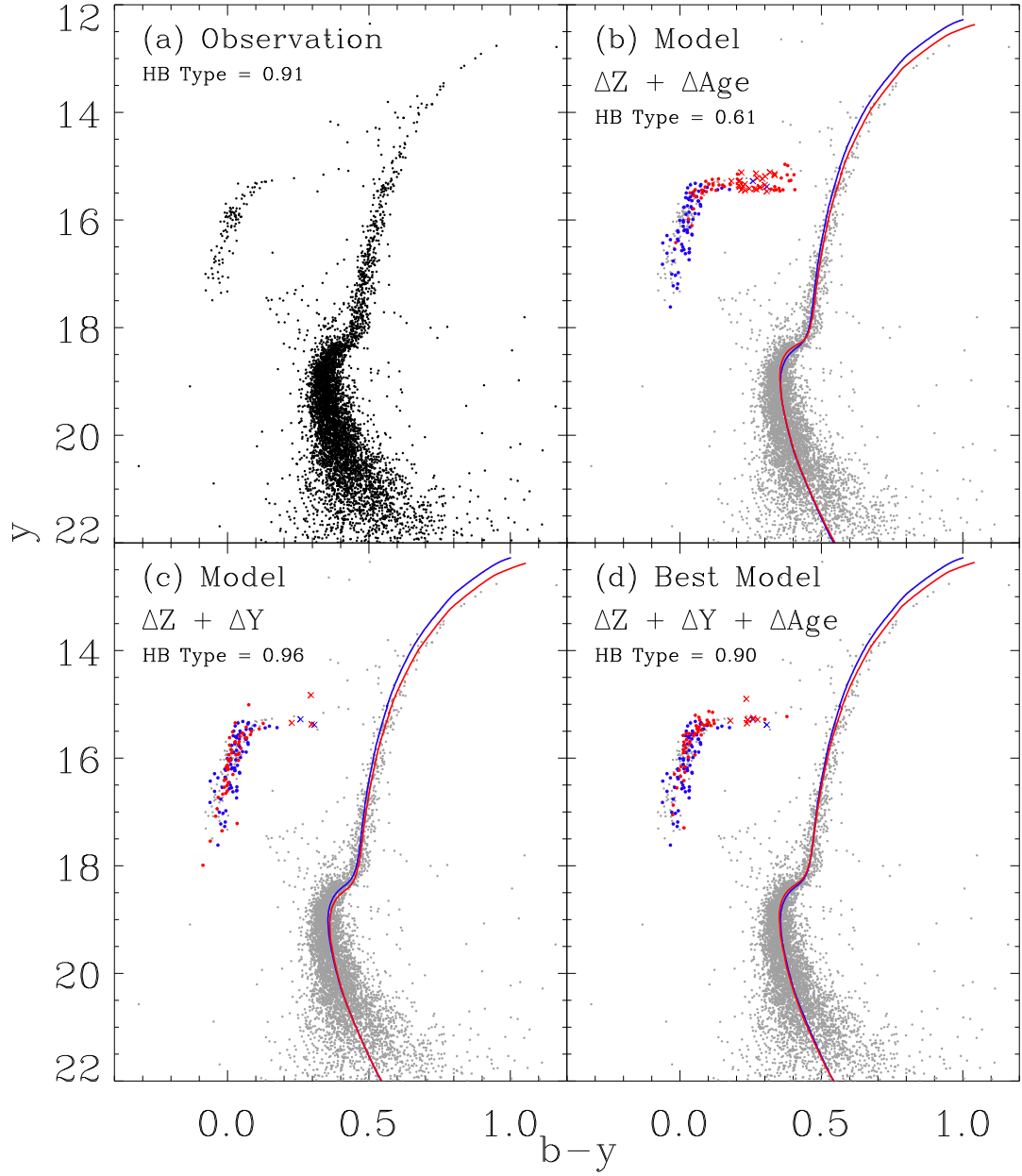}
\caption{Same as Fig. 2, but in $b-y$ vs.\ $y$ plane.
\label{fig3}}
\end{figure*}

\clearpage
\begin{figure*}[ht]
\epsscale{0.9}
\plotone{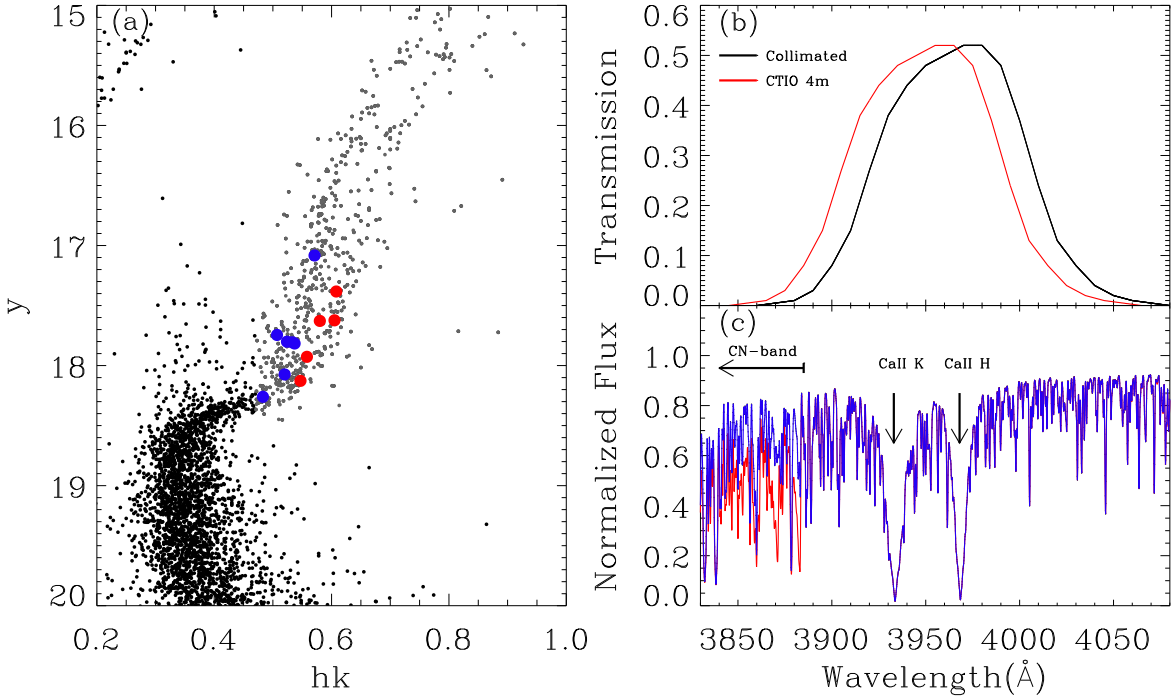}
\caption{(a) Same as Fig. 1b, but zoomed around the SGB and RGB regions in $hk$ vs.\ $y$ CMD.
Denoted by blue and red dots are respectively ``CN normal'' and ``CN strong'' stars from \citet{Kayser08}.
The CN strength seems to trace the calcium abundance.
(b) A comparison of $Ca$ filter transmission functions.
The $Ca$ filter at the prime focus of the CTIO 4m telescope is shifted approximately 15 \AA$ $ to the shorter wavelength compared to the case of perfectly collimated beam.
(c) Synthetic spectra for the CN normal (the blue line) and the CN strong (the red line) RGB stars \citep[see also Supplementary Fig. 1b of ][]{Lee09a}.
Although the CN band starting at $\lambda$~$\approx$ 3885 \AA$ $ lies on the blue tail of $Ca$ filter transmission function, contamination from the CN band appears to be negligible (see text).
\label{fig4}}
\end{figure*}

\clearpage
\begin{deluxetable}{lll}
\footnotesize
\tablecaption{\footnotesize Input Parameters Adopted in Our Best Simulation of NGC 288 \label{tbl-1}}
\tablewidth{0pt}
\tablehead{\colhead{Parameter}\quad \quad \quad \quad \quad \quad&\colhead{Population 1}\quad \quad \quad \quad \quad \quad&\colhead{Population 2}
}
\startdata
Z                              &0.00083          &0.00121          \\
Y                              &0.231            &0.258            \\
$ $[$\alpha$/Fe]               &0.3              &0.3              \\
Age                            &13.7$\pm$0.3~Gyr &12.2$\pm$0.3~Gyr \\
$\eta$\tablenotemark{a}        &0.53             &0.53             \\
$\Delta$$M$\tablenotemark{b}   &0.2179           &0.2094           \\
$\sigma_{M}$\tablenotemark{c}  &0.020            &0.020            \\
Population Ratio               &0.6              &0.4
\enddata
\tablenotetext{a}{\citet{Reimers77} mass-loss parameter.}
\tablenotetext{b}{Mean mass-loss on the RGB ($M_\odot$).}
\tablenotetext{c}{Mass dispersion on the HB ($M_\odot$).}
\end{deluxetable}

\end{document}